\documentclass[12pt]{book}

\usepackage[dvips]{graphicx,color}
\usepackage{makeidx,hori,zon}

\makeauthorindex

\BookTitle{Inflating Horizon of Particle Astrophysics and Cosmology}
\CopyRight{\copyright 2006 by Universal Academy Press, Inc.\ and Yamada Science Foundation}

\begin{document}

\pagenumbering{arabic}
\setcounter{page}{219}

\chapter{
$N$-$z$ Relation and CMB Anisotropies in the Universe with an Oscillating Scalar Field Having a Null Field State}

\author{%
Koichi HIRANO, Kiyoshi KAWABATA, Zen KOMIYA and Hiroshi BUNYA\\
{\it Department of Physics, College of Science,
Tokyo University of Science, 1-3 Kagurazaka, Shinjuku-ku,
Tokyo 162-8601, Japan
}}
%
%
\AuthorContents{K.\ Hirano, K.\ KAWABATA, Z.\ KOMIYA and H.\ BUNYA} 

\AuthorIndex{Hirano}{K.} 
\AuthorIndex{Kawabata}{K.} 
\AuthorIndex{Komiya}{Z.} 
\AuthorIndex{Bunya}{H.} 



\section*{Abstract}
\vspace{-2mm}
We investigate whether or nor it is possible to find a scalar field model or models that are capable of explaining simultaneously the observed $N$-$z$ relation given by the 2dF Galaxy Redshift Survey\cite{Colless}, which still seems to exhibit a spatial periodicity of the galaxy distribution(the 'picket-fence structure'), and the CMB spectrum obtained by the WMAP experiments\cite{Spergel}. It is found that both the observed size of the spatial periodicity and the amplitude of the 2dF $N$-$z$ relation can be fairly well fitted by the theoretical computations based on the scalar field models with $-20\le \xi\le -10$, and $140\le m_{\rm s} \le 160$, where $\xi$ is the gravitational coupling parameter, and $m_{\rm s}$ the normalized mass of the scalar field, respectively. To reproduce the CMB spectrum of the WMAP, we find that it is very crucial to have a null state of the scalar field in the early epochs of evolution of the universe.

\vspace{1mm}

\section{Introduction and Basic Equations}
\vspace{-2mm}
The oscillating scalar field model\cite{Kashino} has several fascinating merits including the fact that it can naturally account for the existence of the spatial periodicity observed in the galaxy number counts $N$ vs. the redshift parameter $z$ relation(the $N$-$z$ relation). 
The evolution equations of the scalar field models of the universe\cite{Kashino} are
\begin{equation}
\frac{\dot{a}}{a}=\frac{6\xi\phi\dot{\phi}+\sqrt{(6\xi\phi\dot{\phi})^2+(1-6\xi\phi^2)\{(\dot{\phi})^2+a^2\frac{H_0^2}{c^2}m_{\rm s}^2\phi^2+\frac{8\pi G}{3c^2}a^2(\rho_r+\rho_m)+\frac{a^2}{3}\Lambda\}}}{1-6\xi\phi^2}
\end{equation}
\begin{equation}
\ddot{\phi}=-2\frac{\dot{a}}{a}\dot{\phi}-6\xi\frac{\ddot{a}}{a}\phi-a^2\frac{H_0^2}{c^2}m_{\rm s}^2\phi ~~~~~~~~~~~~(\phi\equiv\sqrt{\frac{4\pi G}{3c^4}}\psi,~~m_{\rm s} \equiv \frac{mc^2}{\hbar H_0})
\end{equation}
where a dot(~$\dot{}$~) denotes the derivative with respect to the conformal time, $\psi$  the scalar field, $m$ the mass of the scalar field, $\xi$ the gravitational coupling parameter, and the other symbols have their usual meanings.


\vspace{1mm}

\section{$N$-$z$ Relation}
\vspace{-2mm}
In the observational data for the $N$-$z$ relation  obtained through the 2dF Galaxy Redshift Survey\cite{Colless}, we clearly notice that regions(or epochs) of comparatively higher galaxy number counts tend to appear  with a redshift interval $\Delta z$ of about 0.03. We believe that we can account for this so called 'picket-fence structure' by means of our scalar field model rather straightforwardly without recourse to the effect due to 'peculiar velocity'. 
One notable characteristic feature of the oscillating scalar field model universe is that the Hubble parameter oscillates with time.
Since the co-moving volume $V$ depends on the Hubble parameter $H(z)$, the effect of the adopted model universe shows up in the $N$-$z$ relation which sensitively depends on $V$(see Fig.1).
\vspace{-4mm}
\begin{figure}[h]
  \begin{center}
    \begin{minipage}{80mm}
      \begin{center}
    \includegraphics[height=13pc]{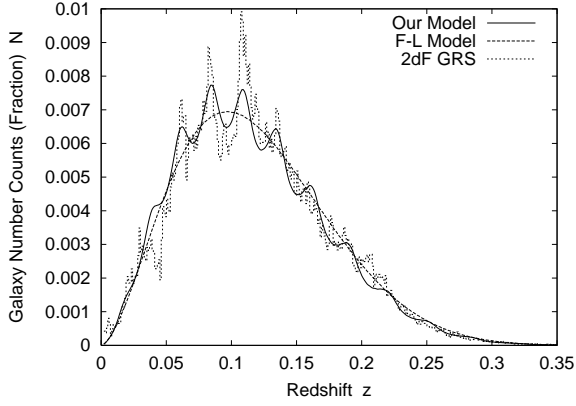}
    \end{center}
    \end{minipage}
    \begin{minipage}{80mm}
      \begin{center}
	\caption{Theoretical and Observed $N$-$z$ Relations:\quad The dotted line is for the 2dF data, the dashed line is for 
	 the F-L model($\Omega_{m,0}$=0.27,~$\Omega_{\Lambda}$=0.73), and  the solid line is for our model with  $\Omega_{m,0}$=0.20, $\Omega_{\Lambda}$=0.50, $h_0$=0.70, $\xi$=$-15$, $m_{\rm s}$=150 and $\phi_0$=0.0005.}
      \end{center}
    \end{minipage}
  \end{center}
\end{figure}

\vspace*{-9mm}

\section{CMB Anisotropies with Oscillating Scalar Field Having a Null Field State}
\vspace{-2mm}
As an important consequence of our analysis, Unless the value of $\phi$ is kept stationary at almost 0 sufficiently long
in the early stage of expansion, we find that we could not reproduce the WMAP observation\cite{Spergel} of the CMB angular spectrum,
 as is shown in Fig.2.
Fig.3 compares the WMAP observation of the CMB angular spectrum with those computed for our model as well as for the flat FL-model.
Note that taking into account the effect of the scalar field fluctuation, which our current model does not, should improve the goodness of fit of our model in the large scale region\cite{Caldwell}. The work along this line is now in progress.
\vspace{-5mm}
\begin{figure}[h]
  \begin{center}
    \begin{minipage}{80mm}
      \begin{center}
      \includegraphics[height=11pc]{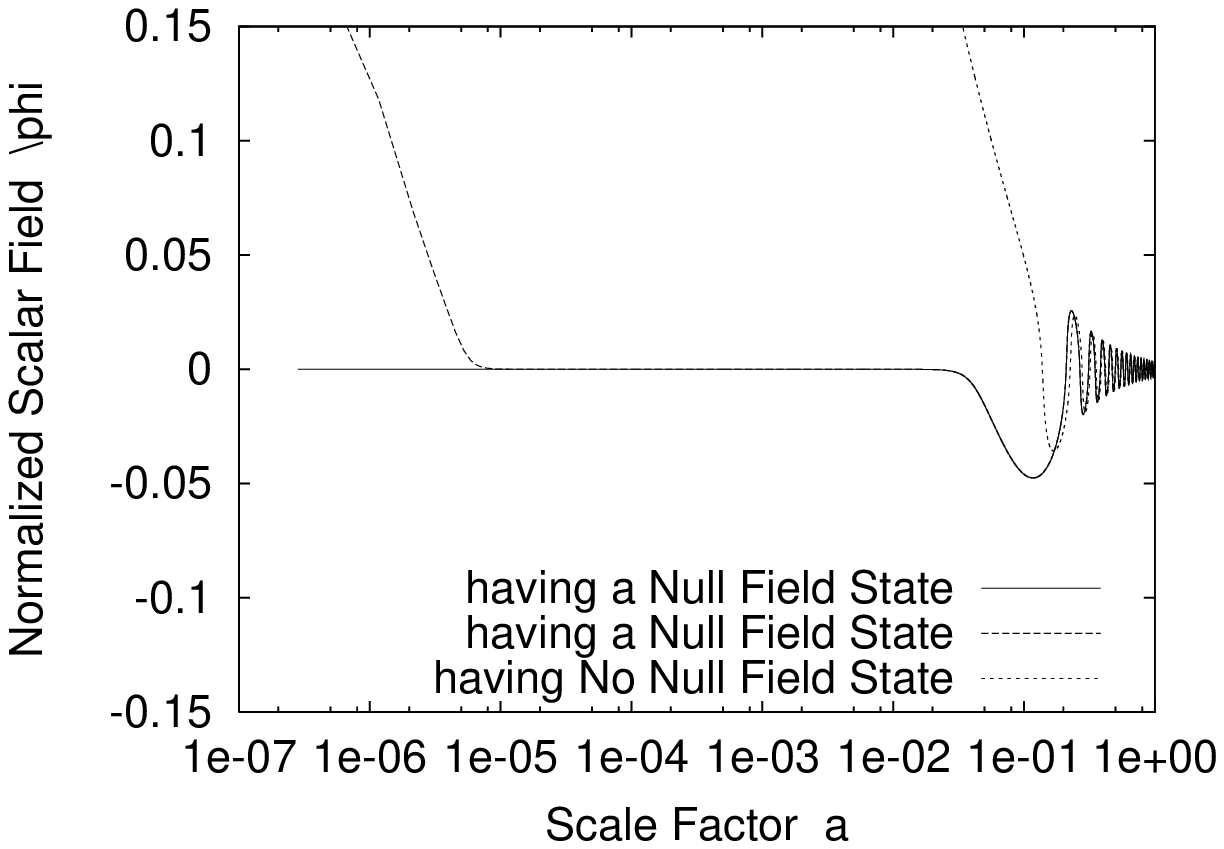}
	\vspace{-2mm}
      \caption{Values of Scalar Field  as Functions of Scale Factor}
      \end{center}
    \end{minipage}
    \hspace{5mm}
    \begin{minipage}{80mm}
      \begin{center}
      \includegraphics[height=10pc]{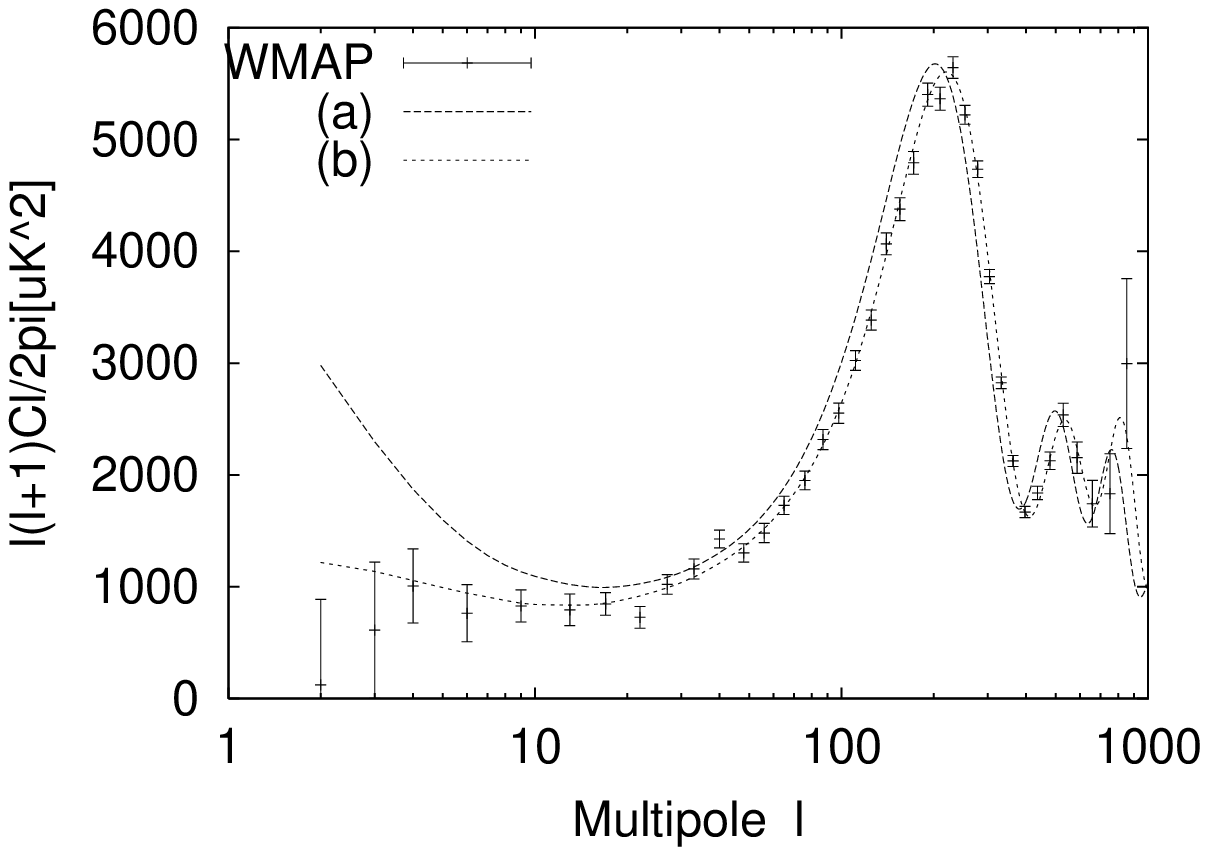}
	\vspace{-2mm}
      \caption{CMB temperature anisotropy spectra
               (a) the $N$-$z$ relation fit model above 
               (b) F-L fit model\cite{Spergel}}
      \end{center}
    \end{minipage}
  \end{center}
\end{figure}

\vspace{-8mm}

\section{Conclusions}
\vspace{-2mm}
In the case of our model, the resulting $N$-$z$ relation is particularly 
sensitive to the assume values of $\xi$ and $m_{\rm s}$.
 It is found that both the observed size of the spatial periodicity and the amplitude of the 2dF $N$-$z$ relation can be fairly well reproduced  by our 
models with $-20\le \xi\le -10$, and $140\le m_{\rm s} \le 160$. 
Furthermore, we have found it essential to suppress the scalar field
 to almost null in the early epochs of evolution of our universe in order to reproduce the WMAP observation.

\end{document}